\documentclass[fleqn,twoside,twocolumn,nofootinbib,showkeys]{revtex4} % Specifies the document class %,unsortedaddress
\usepackage[sec,nopacs,nocpr]{Format}

\usepackage{color,soul}
\usepackage{amsmath,amssymb}
\usepackage[breaklinks=true,colorlinks=true,linkcolor=blue,urlcolor=blue,citecolor=blue]{hyperref}
\numberwithin{equation}{section}
\usepackage[english,ukrainian]{babel}
\usepackage[T1,T2A]{fontenc}
\usepackage[cp1251]{inputenc}
\usepackage{upgreek}

\begin{document}
\title[Temperature dependence of spin pinning and spin-wave dispersion in nanoscopic ferromagnetic waveguides]{Temperature dependence of spin pinning and spin-wave dispersion in nanoscopic ferromagnetic waveguides}
\author{B. Heinz$^*$}
    \affiliation{Fachbereich Physik and Landesforschungszentrum OPTIMAS, Technische Universit\"at Kaiserslautern}
    \address{Erwin-Schr\"odinger-Stra\ss e 56, 67663 Kaiserslautern, Germany}
    \affiliation{Graduate School Materials Science in Mainz}
	\address{Staudingerwerg 9, 55128 Mainz, Germany}
\author{Q. Wang$^*$}
    \affiliation{Fachbereich Physik and Landesforschungszentrum OPTIMAS, Technische Universit\"at Kaiserslautern}
    \address{Erwin-Schr\"odinger-Stra\ss e 56, 67663 Kaiserslautern, Germany}
    \address{Staudingerweg 9, 55128 Mainz, Germany}
\author{R. Verba}
    \affiliation{Institute of Magnetism}
    \address{Vernadskogo blvd. 36-b, 03142 Kyiv, Ukraine}
\author{V.~I.~Vasyuchka}
    \affiliation{Fachbereich Physik and Landesforschungszentrum OPTIMAS, Technische Universit\"at Kaiserslautern}
    \address{Erwin-Schr\"odinger-Stra\ss e 56, 67663 Kaiserslautern, Germany}
\author{M. Kewenig}
\author{P. Pirro}
\author{M. Schneider}
    \affiliation{Fachbereich Physik and Landesforschungszentrum OPTIMAS, Technische Universit\"at Kaiserslautern}
    \address{Erwin-Schr\"odinger-Stra\ss e 56, 67663 Kaiserslautern, Germany}
\author{T.~Meyer}
    \affiliation{Fachbereich Physik and Landesforschungszentrum OPTIMAS, Technische Universit\"at Kaiserslautern}
    \address{Erwin-Schr\"odinger-Stra\ss e 56, 67663 Kaiserslautern, Germany}
    \affiliation{THATec Innovation GmbH}
    \address{Augustaanlage 23, 68165 Mannheim, Germany}
\author{B.~L\"agel}
    \affiliation{Nano Structuring Center, Technische Universit\"at Kaiserslautern}
    \address{Erwin-Schr\"odinger-Stra\ss e 13, 67663 Kaiserslautern, Germany}
\author{C.~Dubs}
    \affiliation{INNOVENT e.V., Technologieentwicklung}
    \address{Pr\"ussingstra\ss e 27B, 07745 Jena, Germany}
\author{T. Br\"acher}
    \affiliation{Fachbereich Physik and Landesforschungszentrum OPTIMAS, Technische Universit\"at Kaiserslautern}
    \address{Erwin-Schr\"odinger-Stra\ss e 56, 67663 Kaiserslautern, Germany}
\author{O.~V.~Dobrovolskiy}
    \affiliation{Physikalisches Institut, Goethe-Universit\"at Frankfurt}
    \address{Max-von-Laue-Str. 1, 60438 Frankfurt am Main, Germany}
\author{A.~V.~Chumak$^{**}$}
    \affiliation{Fachbereich Physik and Landesforschungszentrum OPTIMAS, Technische Universit\"at Kaiserslautern}
    \address{Erwin-Schr\"odinger-Strasse 56, 67663 Kaiserslautern, Germany}
    \affiliation{Nanomagnetism and Magnonics, Faculty of Physics, University of Vienna}
    \address{Boltzmanngasse 5, A-1090 Wien, Austria}
    
\razd{\secviii}

\autorcol{B. Heinz et al.}

\setcounter{page}{1}%

\begin{abstract}
The field of magnonics attracts significant attention due to the possibility of utilizing information coded into the spin-wave phase or amplitude to perform computation operations on the nanoscale. Recently, spin waves were investigated in Yttrium Iron Garnet (YIG) waveguides with widths ranging down to 50 nm and aspect ratios thickness over width approaching unity. A critical width was found, below which the exchange interaction suppresses the dipolar pinning phenomenon and the system becomes unpinned. Here we continue these investigations and analyse the pinning phenomenon and spin-wave dispersions as a function of temperature, thickness and material of choice. Higher order modes, the influence of a finite wavevector along the waveguide and the impact of the pinning phenomenon on the spin-wave lifetime are discussed as well as the influence of a trapezoidal cross section and edge roughness of the waveguides. The presented results are of particular interest for potential applications in magnonic devices and the incipient field of quantum magnonics at cryogenic temperatures.

\end{abstract}


\maketitle

\section{Introduction}

The field of magnonics proposes a promising approach for a novel type of computing systems, in which magnons, the quanta of spin waves, carry the information instead of electrons \cite{Chu15nph,Kru10jpd,Dav15prb,Khi10jpd,Sch16arx,Kra14pcm,Win16nan,Bra18jap,Wan18sca,Zog15iee,Man18nph,Chu19arx}. Since the phase of a spin wave provides an additional degree of freedom efficient computing concepts can be used resulting in a valuable decrease in the footprint of logic units. Moreover, the scalability of magnonic structures down to the nanometer scale and the possibility to operate with spin waves of nanometer wavelength are additional advantages of the magnonics approach. The further miniaturization will, consequently, result in an increase in the frequency of spin waves used in the devices from the currently employed GHz range up to the THz range. In classical magnonics, spin-wave modes in thin films or rather planar waveguides with thickness-to-width aspect ratios $a_\mathrm{r} = h/w \ll 1$ have been utilized. In the case of a waveguide, edge magnetostatic charges arise, which can be accounted for by the introduction of boundary conditions \cite{Iva02apl}. Therefore thin waveguides demonstrate the effect of "dipolar pinning" at the lateral edges, and for its theoretical description the thin strip approximation was developed, in which only pinning of the much-larger-in-amplitude dynamic in-plane magnetization component is taken into account \cite{Rad59pcs,Dam61pcs,Gus02prb,Gus05prb,Gus11mmm,Ari16prb}.

The recent progress in fabrication technology leads to the development of nanoscopic magnetic devices in which the width $w$ and the thickness $h$ become comparable \cite{Bra17prb,Dem15trm,Ciu16apl,Pir14apl,Mru17prb,Hal16acs,Ver13apl,Bra17phr}. The description of such waveguides is beyond the thin strip model of effective pinning, because the scale of nonuniformity of the dynamic dipolar fields, which is described as "effective dipolar boundary conditions", becomes comparable to the waveguide width. Additionally, both, in-plane and out-of-plane dynamic magnetization components, become involved in the effective dipolar pinning, as they become of comparable amplitude. Thus, a more general model should be developed and verified experimentally. In addition, such nanoscopic feature sizes imply that the spin-wave modes bear a strong exchange character, since the widths of the structures are now comparable to the exchange length \cite{Abo13trm}. A proper description of the spin-wave eigenmodes in nanoscopic strips which considers the influence of the exchange interaction, as well as the shape of the structure, was recently performed in \cite{Wan19prl} and is fundamental for the field of magnonics.

Very recently, the fields of quantum magnonics and magnonics at cryogenic temperatures were established. Among the highlights, one could mention the first realization of a coherent coupling between a ferromagnetic magnon and a superconducting qubit \cite{Tab15sci}, the first observation of the interaction between magnons and Abrikosov fluxes in supercondutor-ferromagnet hybrid structures \cite{Dob19nph} and the investigation of the interplay of magnetization dynamics with a microwave waveguide at cryogenic temperatures \cite{Gol19pra}. Thus, the understanding of the influence of temperature on spin pinning conditions and on spin-wave dispersions in nano-structures is of high demand.

In this Article, we continue the investigation carried out in Phys. Rev. Lett. \textbf{122}, 247202 (2019). The evolution of the frequencies and profiles of the spin-wave modes in Yttrium Iron Garnet (YIG) waveguides with a thickness of $39$\,nm and widths ranging down to $50$\,nm are discussed in detail. The phenomenon of unpinning and the underlying theory as well as the experimental proof are outlined. A throughout discussion of the effective width and the critical width, at which the system becomes unpinned, in dependency of the thickness and the material of choice is presented. Moreover, the temperature dependency is analysed theoretically. Higher order modes up to $n=2$, the influence of a finite wavevector along the waveguide and the impact of the pinning phenomenon on the spin wave lifetime are discussed. To account for the imperfections of a real system, the influence of a trapezoidal cross section and edge roughness on the effective width and the critical width are investigated.

\begin{figure}[tbh!]
    \vskip1mm
    \includegraphics[width=\column]{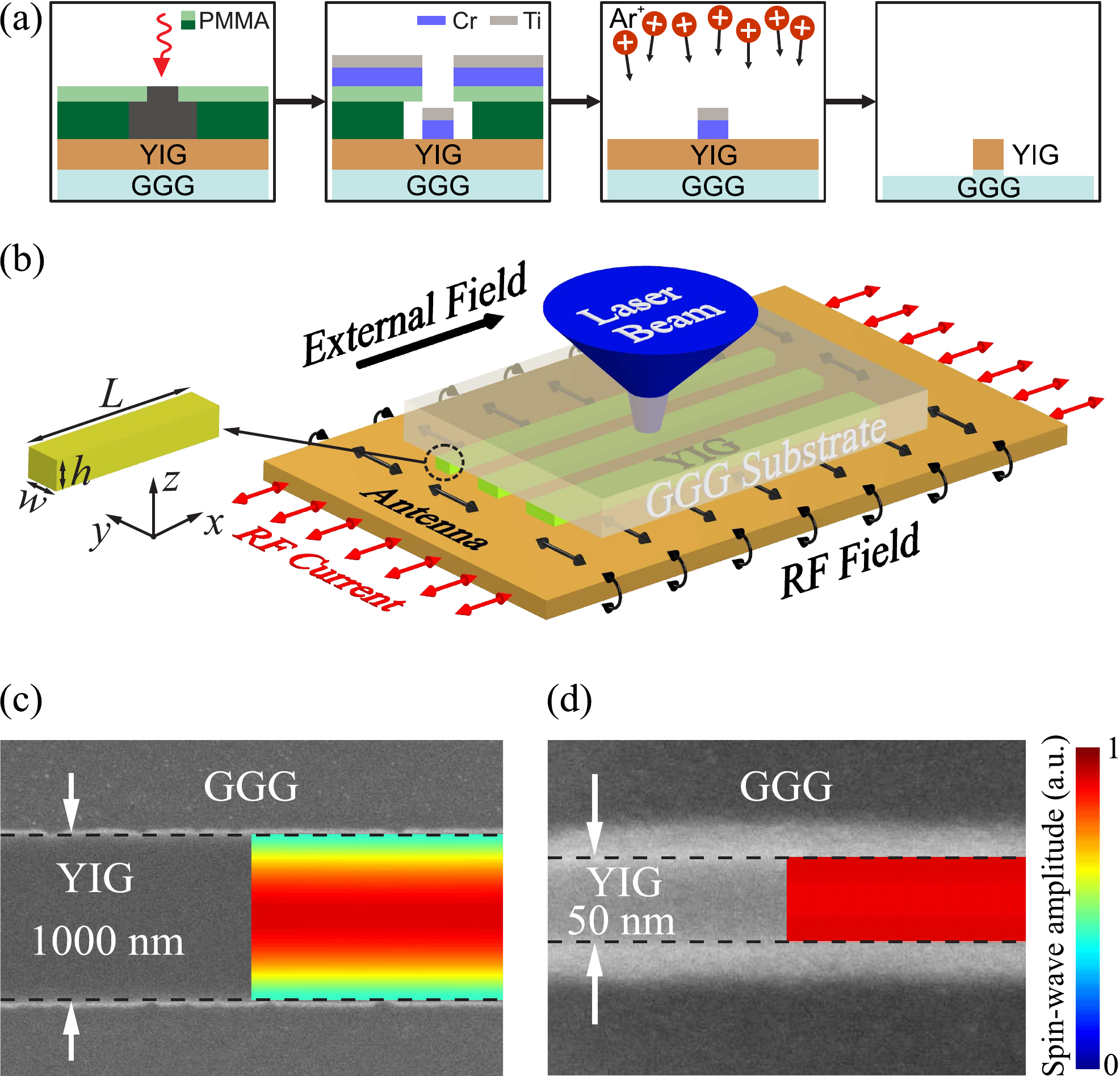}
    \vskip-3mm\caption{(a) Schematically depicted main steps in the nanostructuring process. (b) Sketch of the sample and the experimental configuration: a set of YIG waveguides is placed on a microstrip line to excite the quasi-FMR in the waveguides. BLS spectroscopy is used to measure the local spin-wave dynamics. (c) and (d) SEM micrograph of a $1\,\mu$m and a $50$-nm wide YIG waveguide of $39$-nm thickness. The color code shows the simulated amplitudes of the fundamental mode at quasi-ferromagnetic resonance, i.e., $k_x = 0$, in the waveguides. The mode in the $50\,n$m waveguide is almost uniform across the width of the waveguide evidencing the unpinning directly. \cite{Wan19prl}}
    \label{fig1}
\end{figure}

\section{Methodology}

\subsection{Sample fabrication}

A $39$\,nm thick Yttrium Iron Garnet (YIG) film has been grown on a 1 inch (111) $500\,\mu$m thick Gadolinium Gallium Garnet (GGG) substrate by liquid phase epitaxy from PbO-B$_2$O$_3$ based high-temperature solutions at $860^\circ\text{C}$ using the isothermal dipping method (see e.g. Ref. \cite{Dub17jpd}). A pure Y$_3$Fe$_5$O$_{12}$ film with a smooth surface was obtained by rotating the substrate horizontally with a rotation rate of $100$\,rpm. The saturation magnetization of the YIG film is $1.37 \times 10^5$\,A/m and its Gilbert damping is $\alpha = 6.41 \times 10^{-4}$, as it was extracted by ferromagnetic resonance spectroscopy \cite{Mak15phy}.

The nanostructures were fabricated by utilising a hard mask ion milling procedure. The key steps in the fabrication process are shown in Fig.  \ref{fig1}(a). First a double layer of polymethyl methacrylate (PMMA) was spin coated on the YIG film and a Chromium/Titanium hard mask was fabricated using electron beam lithography and electron beam evaporation. This hard mask acts as a protective layer in a successive Ar$^+$ ion milling step. In a final step, any residual Chromium is removed using an acid that YIG is inert to.

\subsection{Microfocused Brillouin Light Scattering (BLS) spectroscopy measurements}
BLS spectroscopy is a unique technique for measuring the spin-wave intensities in frequency, space, and time domains. It is based on the inelastic scattering of an incident laser beam from a magnetic material. In our measurements, a laser beam of $457$\,nm wavelength and a power of $1.8$\,mW is focused through the transparent GGG substrate on the center of the respective individual waveguide using a $100\times$ microscope objective with a large numerical aperture ($\mathrm{NA}=0.85$). The effective spot-size is $350$\,nm. The scattered light was collected and guided into a six-pass Fabry-P\'erot interferometer to analyse the frequency shift.

\subsection{Numerical simulations}
The micromagnetic simulations of the space- and time dependent magnetization dynamics were performed by the GPU-accelerated simulation program Mumax3 using a finite-difference discretisation \cite{Van14aip}. The structure is schematically shown in Fig. \ref{fig1}(b). The following material parameters were used in the simulations: the saturation magnetization $M_\mathrm{s} = 1.37\times10^5$\,A/m and the Gilbert damping   $\alpha = 6.41\times10^{-4}$ were extracted by ferromagnetic resonance spectroscopy measurements of the plain film before patterning \cite{Mak15phe}. A gyromagnetic ratio of $\gamma = 175.86$\,rad/(ns$\cdot$T) and an exchange constant of $A = 3.5$\,pJ/m for a standard YIG film were assumed. An external field $B = 108.9$\,mT is applied along the waveguide long axis. Three steps were performed to calculate the spin-wave dispersion curve: (i) The external field was applied along the waveguide, and the magnetization was allowed to relax into a stationary state (ground state). (ii) A sinc field pulse $b_y=b_0\mathrm{sinc}(2 \pi f_ct)$, with oscillation field $b_0 = 1$\,mT and cut-off frequency $f_c = 10$\,GHz, was used to excite a wide range of spin waves. (iii) The spin-wave dispersion relations were obtained by performing the two-dimensional Fast Fourier Transformation of the time- and space-dependent data. Furthermore, the spin-wave width profiles were extracted from the $m_z$ component across the width of the waveguides using a single frequency excitation.

\subsection{Quasi-analytical theoretical model}
In order to accurately describe the spin-wave characteristics in nanoscopic longitudinally magnetized waveguides, a more general semi-analytical theory is provided which goes beyond the thin strip approximation \cite{Wan19prl}. Here, we assume a uniform spin wave mode profile across the waveguide thickness (no $z$-dependency), which is valid for the fundamental thickness mode in thin waveguides. In a general case the z-dependency of any higher thickness mode can be included in a similar way as shown here. Also, please note that the theory is not applicable in transversely magnetized waveguides due to their more involved internal field landscape \cite{Bra17prb}. The lateral spin-wave mode profile $\mathbf{m}_{k_x}(y)$ and frequency can be found as solutions of the linearized Landau-Lifshitz equation \cite{Ver12prb,Ver13ujp}
\begin{equation}
\label{e1}
    -i\omega_{k_x} \mathbf{m}_{k_x}(y) =
    \boldsymbol{\upmu}\times\left(\hat\Omega_{k_x} \cdot \mathbf{m}_{k_x}(y)\right)
\end{equation}
with appropriate exchange boundary conditions, which take into account the surface anisotropy at the edges. Here, $ \boldsymbol{\upmu}$ is the unit vector in the static magnetization direction and $\hat\Omega_{k_x}$ is a tensorial Hamilton operator, which is given by
\begin{equation}
\label{e2}
\begin{array}{lll}
    \hat\Omega_{k_x}\cdot \mathbf{m}_{k_x}(y) =
    \left(\omega_\mathrm{H} + \omega_\mathrm{M}\lambda^2\left(k^2_x - \displaystyle\frac{d^2}{dy^2}\right)\right)\mathbf{m}_{k_x}(y)\\ +
    \omega_\mathrm{M}\int\hat{\mathbf{G}}_{k_x}(y-y^\prime)\cdot \mathbf{m}_{k_x}(y^\prime)dy^\prime.
\end{array}
\end{equation}

Here, $\omega_\mathrm{H} = \gamma B$, $B$ is the static internal magnetic field that is considered to be equal to the external field due to the negligible demagnetization along the $x$-direction, $\omega_\mathrm{M} = \gamma\mu_0 M_\mathrm{s}$, $\gamma$ is the gyromagnetic ratio and $\hat{\mathbf{G}}_{k_x}$ is the Green's function (see next sub-section).

A numerical solution of Eq. (\ref{e1}) gives both, the spin-wave profile $\mathbf{m}_{k_x}$ and frequency $\omega_{k_x}$. In the following, we will regard the ouf-of-plane component $m_z(y)$ to show the mode profiles representatively. In the past, it was demonstrated that in microscopic waveguides, the fundamental mode is well fitted by the function $m_z(y)=A_0\cos(\pi y/w_\mathrm{eff})$ with the amplitude $A_0$ and the effective width $w_\mathrm{eff}$ \cite{Gus02prb,Gus05prb}. This mode, as well as the higher modes, are referred to as ``partially pinned''. Pinning hereby refers to the fact that the amplitude of the mode at the edges of the waveguide is reduced. In that case, the effective width $w_\mathrm{eff}$ determines where the amplitude of the modes would vanish outside the waveguide \cite{Wan18sca,Gus02prb,Bra17phr}. With this effective width, the spin-wave dispersion relation can also be calculated by the analytical formula \cite{Wan18sca}:
\begin{equation}
\label{e3}
\begin{array}{lll}
            \omega_0(k_x) = \\
            \\
            \sqrt{(\omega_\mathrm{H} + \omega_\mathrm{M}(\lambda^2 K^2 + F_{k_x}^{yy})) (\omega_\mathrm{H} + \omega_\mathrm{M}(\lambda^2 K^2 + F_{k_x}^{zz}))},
\end{array}
\end{equation}
where $K = \sqrt{k_x^2 + \kappa^2}$ and $\kappa = \pi/w_\mathrm{eff}$. The tensor $\hat{F}_{k_x}= \frac{1}{2\pi}\displaystyle\int_{\substack{-\infty}}^{\substack{\infty}}\frac{|\sigma_k|^2}{\tilde w}\hat{\mathbf{N}}_k dk_y$ accounts for the dynamic magnetization, $\sigma_k = \displaystyle\int_{\substack{-w/2}}^{\substack{w/2}} m(y)e^{-ik_y y} dy$ is the Fourier-transform of the spin-wave profile across the width of the waveguide and $\tilde w = \displaystyle\int_{\substack{-w/2}}^{\substack{w/2}} m(y)^2 dy$ is the normalization of the mode profile $m(y)$.

\subsection{Numerical solution of the eigenproblem}

In this sub-section we discuss the details of the numerical solution of the eigenproblem. The eigenproblem Eq. (\ref{e1}) should be solved with proper boundary conditions at the lateral edges of the waveguide. We use a complete description of the dipolar interaction via Green's functions:
\begin{equation}
\label{s1}
    \hat{\mathbf{G}}_{k_x}(y) = \frac{1}{2\pi}\int_{-\infty}^{\infty}\hat{\mathbf{N}}_k e^{ik_y y} dk_y.
\end{equation}
Here,
\begin{equation}
\label{s2}
    \hat{\mathbf{N}}_{k} =
     \left(
     \begin{array}{ccc}
              \frac{k_x^2}{k^2}f(kh)    &   \frac{k_x k_y}{k^2}f(kh)        &       0        \\
              \frac{k_x k_y}{k^2}f(kh)      &   \frac{k_y^2}{k^2}f(kh)          &       0         \\
              0                         &           0                       &       1-f(kh)
     \end{array}
     \right),
\end{equation}
where $f(kh)= 1 - (1 - \exp(-kh))/kh$, $k = \sqrt{k_x^2 + k_y^2}$ and it is assumed that the waveguides are infinitely long.

The boundary condition (\ref{s3}) only accounts for the exchange interaction and surface anisotropy (if any) and reads \cite{Gur96boo}
\begin{equation}
\label{s3}
    \mathbf{m}\times(\mu_0M_\mathrm{s} \lambda^2 \frac{\partial \mathbf{m}}{\partial \mathbf{n}} - \nabla_{\mathbf{M}} E_\mathrm{a}) = 0,
\end{equation}
where $\mathbf{n}$ is the unit vector defining the inward normal direction to the waveguide edge, and $E_\mathrm{a}(\mathbf{m})$ is the energy density of the surface anisotropy. In the studied case of a waveguide magnetized along its long axis, the conditions (\ref{s3}) for the dynamic magnetization components can be simplified to
\begin{equation}
\label{s4}
    \pm\frac{\partial m_y}{\partial y} + dm_y|_{y = \pm w/2} = 0,\qquad
    \frac{\partial m_z}{\partial y}|_{y = \pm w/2} = 0,
\end{equation}
where $d = -2 K_\mathrm{s}/(m_0M_\mathrm{s}^2 \lambda^2)$ is the pinning parameter \cite{Gus11mmm} and $K_\mathrm{s}$ is the surface anisotropy constant at the waveguide lateral edges. More complex cases like, e.g., diffusive interfaces can be considered in the same manner \cite{Kru14pcm}.

For the numerical solution of Eq. (\ref{e1}) it is convenient to use finite element methods and to discretize the waveguide into $n$ elements of the width $\Delta w = w/n$, where $w$ is the width of the waveguide. The discretization step should be at least several times smaller than the waveguide thickness and the spin-wave wavelength $2\pi/k_x$ for a proper description of the magneto-dipolar fields. The discretization transforms Eq. (\ref{e1}) into a system of linear equations for magnetizations $\mathbf{m}_j$ , $j = 1,2,3,\cdots n$
\begin{equation}
\label{s5}
\begin{array}{lll}
\displaystyle
    i \boldsymbol{\upmu}\times ((\omega_\mathrm{M} + \omega_\mathrm{M} \lambda^2 k^2_x)\mathbf{m}_j\\
    - \omega_\mathrm{M}\lambda^2\frac{\mathbf{m}_{j-1} - 2\mathbf{m}_j + \mathbf{m}_{j+1}}{\Delta w^2} \\
    + \omega_\mathrm{M}\sum_{j^\prime=1}^n\hat{\mathbf{G}}_{j-j^\prime}\cdot\mathbf{m}_{j^{\prime}} ) = \omega\mathbf{m}_j
\end{array}
\end{equation}
where dipolar interaction between the discretized elements is described by
\begin{equation}
\label{s6}
\begin{array}{lll}
    \hat{\mathbf{G}}_{k_x, j}(y) = \\
    \\
     \frac{1}{\Delta w} \int_{-\Delta w/2}^{\Delta w/2} dy
    \int_{-\Delta w/2}^{\Delta w/2} dy^\prime \hat{\mathbf{G}}_{k_x}(y - y^\prime - j\Delta w).
\end{array}
\end{equation}
The direct use of Eq. (\ref{s6}) is complicated since the Green's function $\hat{\mathbf{G}}_{k_x}(y)$ is an integral itself. Using the Fourier transform it can be derived as
\begin{equation}
\label{s7}
   \hat{\mathbf{G}}_{k_x, j}(y) =
    \frac{\Delta w}{2\pi}\int\mathrm{sinc}(k_y \Delta w/2)
   \hat{\mathbf{N}}_k e^{ik_y j\Delta w} dk_y,
\end{equation}
with $\mathrm{sinc}(x)=\frac{\mathrm{sin}(x)}{x}$. This can be easily calculated, especially using fast Fourier transform. Equation (\ref{s5}) is, in fact, a $2n$-dimensional linear algebraic eigenproblem (since $\mathbf{m}_j$ is a 2-component vector), which is solved by standard methods. The values $\mathbf{m}_0$ and $\mathbf{m}_{n+1}$ in Eq. (\ref{s5}) are determined from the boundary conditions (\ref{s4}). In particular, for negligible anisotropy at the waveguide edges one should set $\mathbf{m}_0 = \mathbf{m}_1$ and $\mathbf{m}_{n+1} = \mathbf{m}_{n}$.

\section{Results and discussions}

\subsection{Original experimental findings}

In these studies, we consider rectangular magnetic waveguides as shown schematically in Fig. \ref{fig1}(b). In the experiment, a spin-wave mode is excited by a stripline that provides a homogeneous excitation field over the sample containing various waveguides etched from a $h = 39$\,nm thick YIG film. The widths of the waveguides range from $w = 50$\,nm to $w = 1\,\mu$m and the length is $60\,\mu$m. The waveguides are uniformly magnetized along their long axis by an external field $B$ ($x$-direction). Figure \ref{fig1}(c) and \ref{fig1}(d) show scanning electron microscopy (SEM) micrographs of the largest and the narrowest waveguide studied in the experiment. The intensity of the magnetization precession is measured by microfocused BLS spectroscopy \cite{Seb15fip} (see Methodology section) as shown in Fig. \ref{fig1}(b). Black and red lines in Fig. \ref{fig3}(a) show the frequency spectra for a $1\,\mu$m and a $50$-nm wide waveguide, respectively. No standing modes across the thickness were observed in our experiment, as their frequencies lie higher than $20$\,GHz due to the small thickness. The quasi-FMR frequency is $5.007$\,GHz for the $1\,\mu$m wide waveguide. This frequency is comparable to $5.029$\,GHz, the value predicted by the classical theoretical model using the thin strip approximation \cite{Gus02prb,Gus05prb,Gus11mmm}. In contrast, the quasi-FMR frequency is $5.35$\,GHz for a $50$\,nm wide waveguide which is much smaller than the value of $7.687$\,GHz predicted by the same model. The reason is that the thin strip approximation overestimates the effect of dipolar pinning in waveguides with aspect ratio either $a_\mathrm{r} =1$ or close to one, for which the nonuniformity of the dynamic dipolar fields is not well-localized at the waveguide edges. Additionally, in such nanoscale waveguides, the dynamic magnetization components become of the same order of magnitude and both affect the effective mode pinning, in contrast to thin waveguides, in which the in-plane magnetization component is dominant.

\subsection{Spin pinning in nano-structures}

In the following, the experiment is compared to the theory and to micromagnetic simulations.
\begin{figure}[tbh!]
    \vskip1mm
    \includegraphics[width=\column]{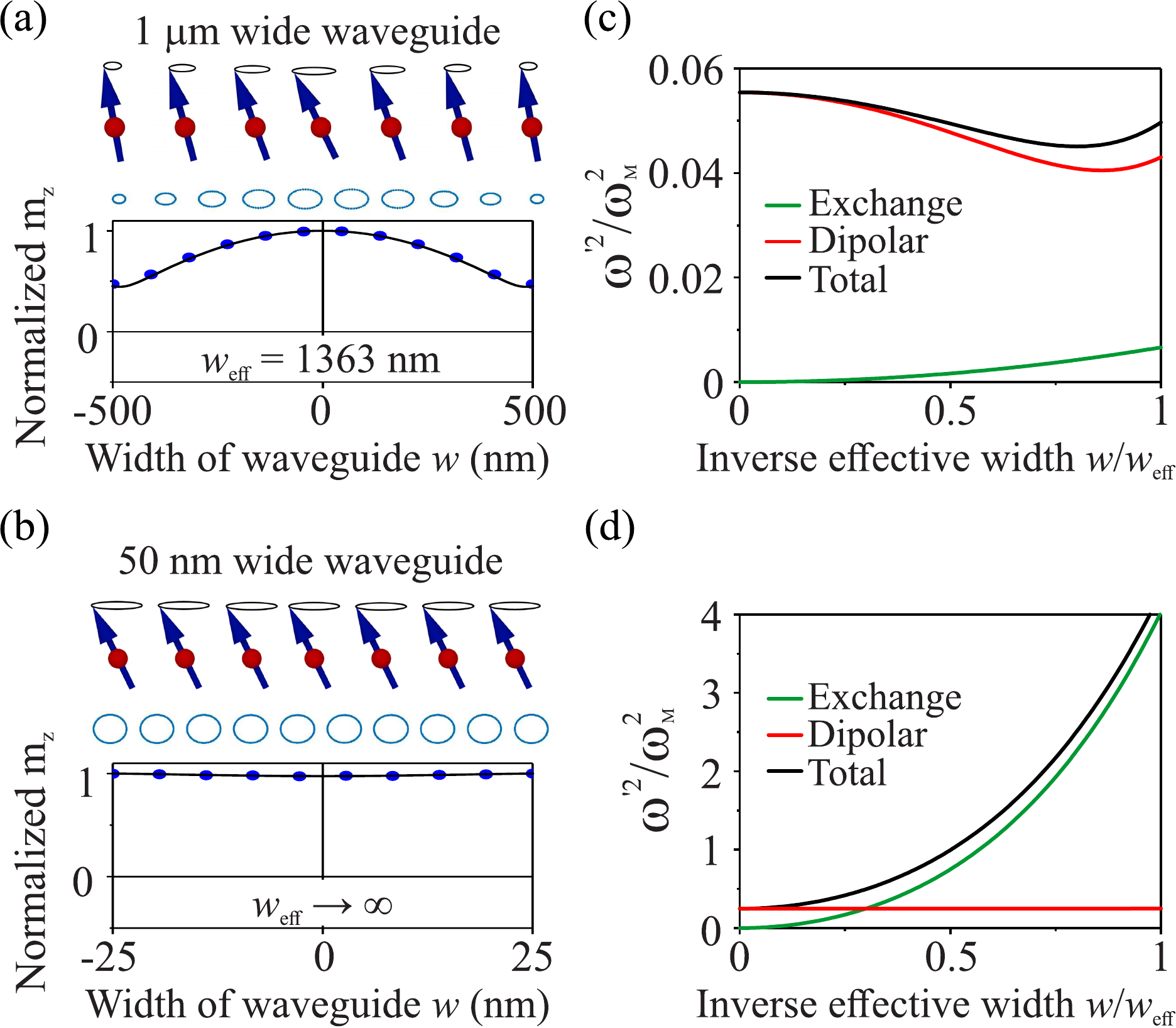}
    \vskip-3mm\caption{Schematic of the precessing spins and simulated precession trajectories (ellipses in the second panel) and spin-wave profile $m_z(y)$ of the quasi-FMR. The profiles have been obtained by micromagnetic simulations (blue dots) and by the quasi-analytical approach (black lines) for a (a) $1\,\mu$m and a (b) $50$ nm wide waveguide. (c), (d): Corresponding normalized square of the spin-wave eigenfrequency $\omega^{\prime2}/\omega_\mathrm{M}^2$ as a function of $w/w_\mathrm{eff}$ and the relative dipolar and exchange contributions. \cite{Wan19prl}}
    \label{fig2}
\end{figure}

The bottom panels of Fig. \ref{fig2}(a) and \ref{fig2}(b) show the spin-wave mode profile of the fundamental mode for $k_x = 0$, which corresponds to the quasi-FMR, in a $1\,\mu$m (a) and $50$\,nm (b) wide waveguide which have been obtained by micromagnetic simulations (blue dots) and by solving Eq. (\ref{e1}) numerically (black lines) (higher width modes are discussed in the next sections). The top panels illustrate the mode profile and the local precession amplitude in the waveguide. As it can be seen, the two waveguides feature quite different profiles of their fundamental modes: in the $1\,\mu$m wide waveguide, the spins are partially pinned and the amplitude at the edges of the waveguide is reduced compared to the maximal value of $m_z =1$. This still resembles the cosine-like profile of the lowest width mode $n = 0$ that has been well established in investigations of spin-wave dynamics in waveguides on the micron scale \cite{Pir14apl,Bra17phr,Jun15jap} and that can be well-described by the simple introduction of a finite effective width $w_\mathrm{eff} > w$ ($w_\mathrm{eff} = w$ for the case of full pinning). In contrast, the spins at the edges of the narrow waveguide are completely unpinned and the amplitude of the dynamic magnetization $m_z$ of the lowest mode $n = 0$ is almost constant across the width of the waveguide, resulting in $w_\mathrm{eff} \rightarrow\infty$.

\begin{figure*}[tbh!]
    \vskip1mm
    \includegraphics[width=0.85\textwidth]{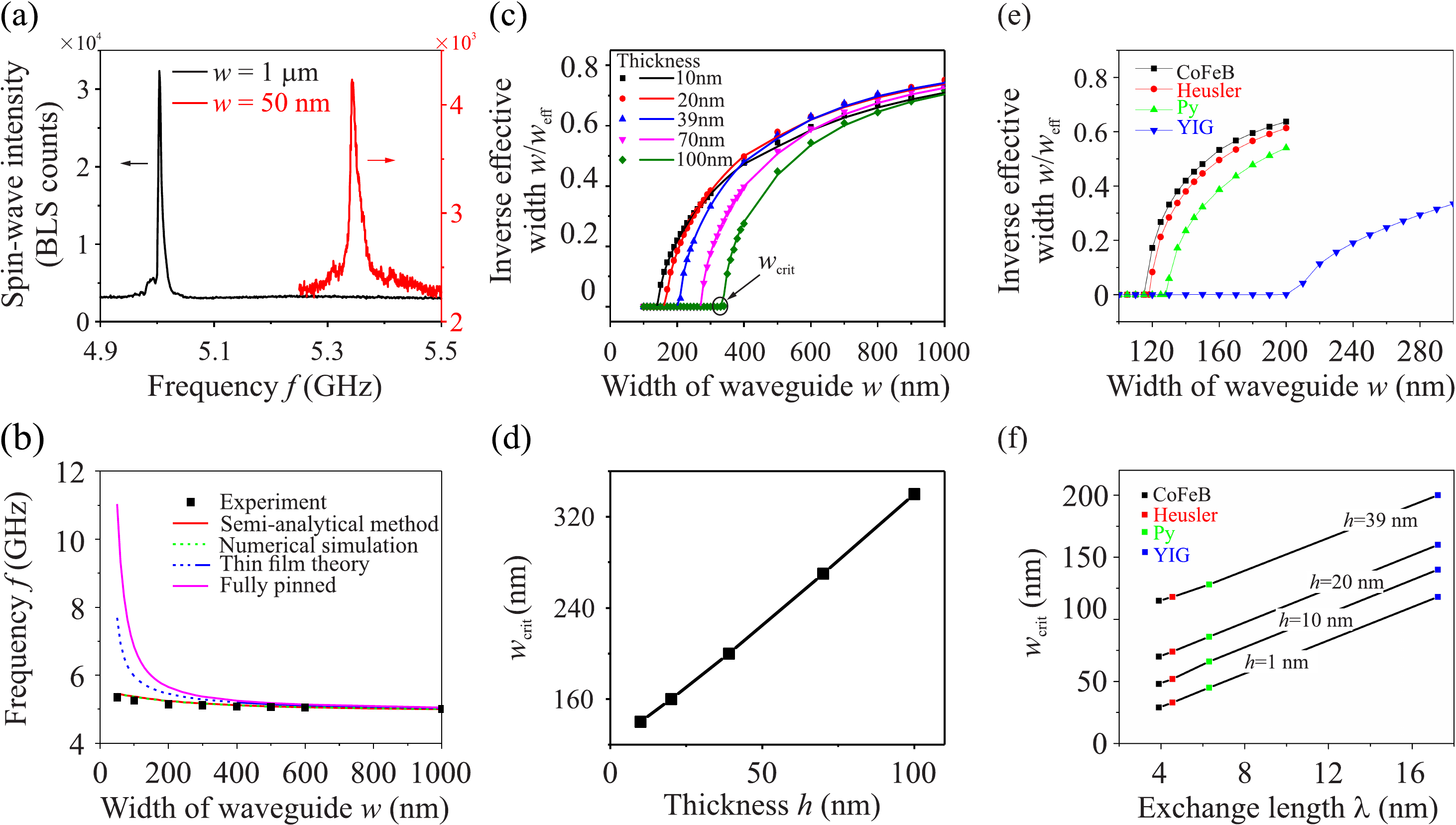}
    \vskip-3mm\caption{(a) Frequency spectra for $1\,\mu$m and $50$\,nm wide waveguides measured for a respective microwave power of $6$\,dBm and $15$\,dBm. (b) Experimentally determined resonance frequencies (black squares) together with theoretical predictions and micromagnetic simulations. (c) Inverse effective width $w/w_\mathrm{eff}$ as a function of the waveguide width. (d) The critical width ($w_\mathrm{crit}$) as a function of thickness $h$. (e) Inverse effective width $w/w_\mathrm{eff}$ as a function of waveguide width for different materials at fixed thickness of 39 nm. (f) The critical width $w_\mathrm{crit}$ as a function of exchange length $\lambda$ for different thicknesses. (a) - (c) \cite{Wan19prl}}
    \label{fig3}
\end{figure*}

To understand the nature of this depinning, it is instructive to consider the spin-wave energy as a function of the geometric width of the waveguide normalized by the effective width $w/w_\mathrm{eff}$. This ratio corresponds to some kind of pinning parameter taking values in between 1 for the fully pinned case and 0 for the fully unpinned case. According to the Ritz’s variational principle, the profiles of the spin wave modes correspond to the respective minima of the spin wave frequency (energy) functional. Since only one minimization parameter – $w_\mathrm{eff}$ – is used, the minimization as a function of $w/w_\mathrm{eff}$ yields only the approximate spin wave profiles. Nevertheless, this is sufficient for the qualitative understanding. To illustrate this, Figs. \ref{fig2}(c) and \ref{fig2}(d) show the normalized square of the spin-wave eigenfrequencies $\omega^{\prime2}/\omega_\mathrm{M}^2$ for the two different widths as a function of $w/w_\mathrm{eff}$. Here, $\omega^{\prime2}$ refers to a frequency square, not taking into account the Zeeman contribution $(\omega^2_\mathrm{H} + \omega_\mathrm{H}\omega_\mathrm{M})$, which only leads to an offset in frequency. The minimum of $\omega^{\prime2}$  is equivalent to the solution with the lowest energy corresponding to the effective width $w_\mathrm{eff}$. In addition to the total $\omega^{\prime2}$ (black), also the individual contributions from the dipolar term (red) and the exchange term (blue) are shown, which can only be separated conveniently from each other if the square of Eq. (\ref{e3}) is considered for $k_x = 0$. The dipolar contribution is non-monotonous and features a minimum at a finite effective width $w_\mathrm{eff}$, which can clearly be observed for $w = 1\,\mu$m. The appearance of this minimum, which leads to the effect known as ``effective dipolar pinning'' \cite{Gus05prb,Gus11mmm}, is a result of the interplay of two tendencies: (i) an increase of the volume contribution with increasing $w/w_{\mathrm{eff}}$, as for common Damon-Eshbach spin waves, and (ii) a decrease of the edge contribution when the spin-wave amplitude at the edges vanishes ($w/w_{\mathrm{eff}} > 1$). This minimum is also present in the case of a $50$\,nm wide waveguide (red line), even though this is hardly perceivable in Fig. \ref{fig2}(d) due to the scale. In contrast, the exchange leads to a monotonous increase of the frequency as a function of $w/w_\mathrm{eff}$, which is minimal for the unpinned case, i.e., $w/w_\mathrm{eff} = 0$ implying $w_\mathrm{eff} \rightarrow\infty$, when all spins are parallel. In the case of the $50$\,nm waveguide, the smaller width and the corresponding much larger quantized wavenumber in the case of pinned spins would lead to a much larger exchange contribution than this is the case for the $1\,\mu$m wide waveguide (please note the vertical scales). Consequently, the system avoids pinning and the solution with lowest energy is situated at $w/w_\mathrm{eff} = 0$. In contrast, in the $1\,\mu$m wide waveguide, the interplay of dipolar and exchange energy implies that the energy is minimized at a finite $w/w_\mathrm{eff}$.

\begin{figure*}[tbh!]
    \vskip1mm
    \includegraphics[width=0.8\textwidth]{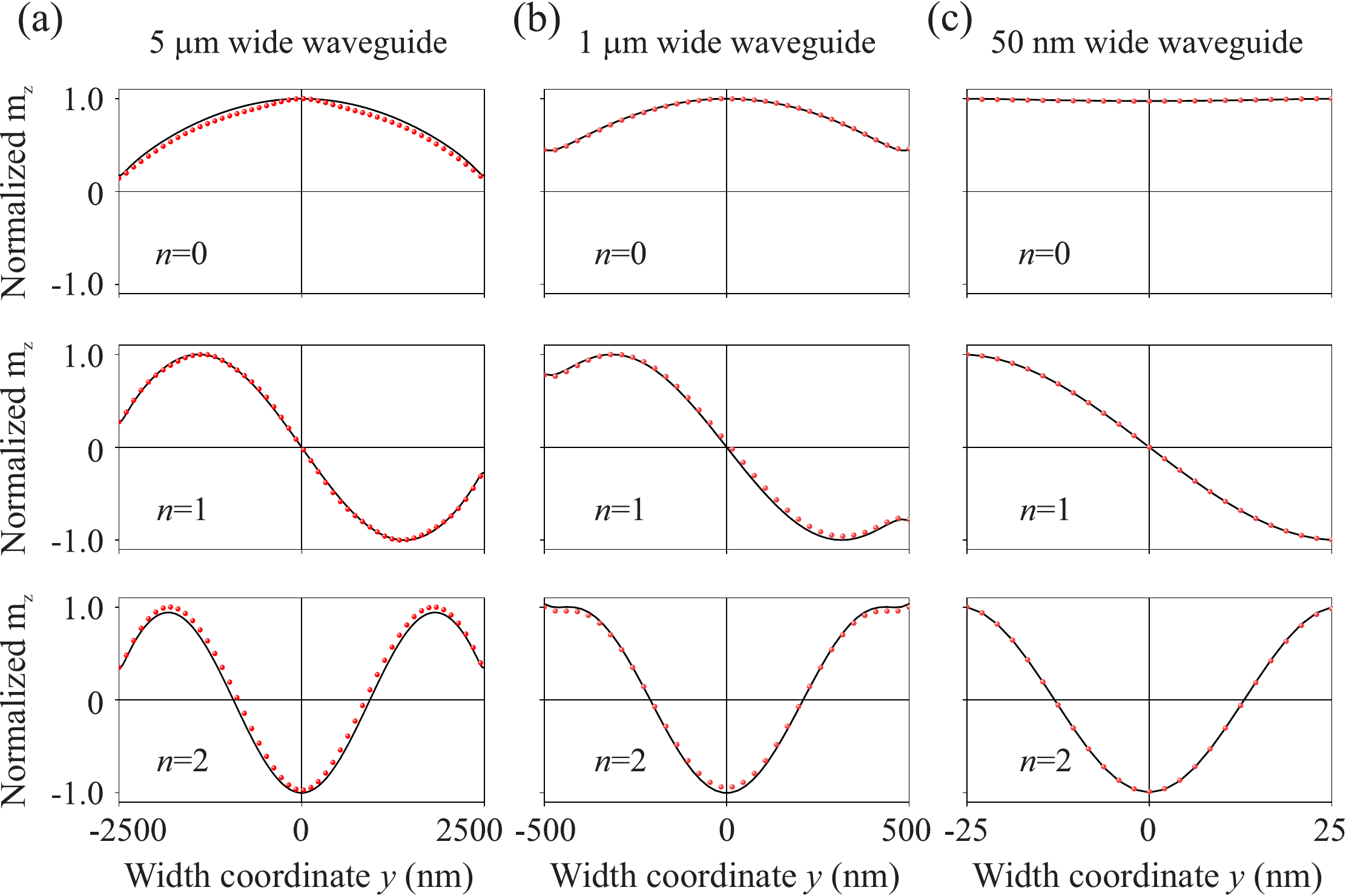}
    \vskip-3mm\caption{The spin-wave profile representatively depicted using the $m_z$ component of the dynamic magnetization for the three lowest width modes obtained by micromagnetic simulation (black solid lines) and numerical calculation (red dots) for (a) $5\,\mu$m, (b) $1\,\mu$m and (c) $50$\,nm wide waveguides.}
    \label{fig4}
\end{figure*}

\subsection{The dependence of the spin-wave frequency on the spin pinning and the critical width of the exchange unpinning}

As it is evident from Figs. \ref{fig2}(c) and \ref{fig2}(d), the pinning and the corresponding effective width have a large influence on the spin-wave frequency. This allows for an experimental verification of the presented theory, since the frequency of partially pinned spin-wave modes would be significantly higher than in the unpinned case. Black squares in Fig. \ref{fig3}(b) summarize the dependence of the frequency of the quasi-FMR on the width of the YIG waveguide. The magenta line shows the expected frequencies assuming pinned spins, the blue (dashed) line gives the resonance frequencies extrapolating the formula conventionally used for micron-sized waveguides \cite{Kal86jpc} to the nanoscopic scenario, and the red line gives the result of the theory presented here, together with simulation results (green dashed line). As it can be seen, the experimentally observed frequencies can be well reproduced if the real pinning conditions are taken into account.

As has been discussed alongside with Fig. \ref{fig2}, the unpinning occurs when the exchange interaction contribution becomes so large that it compensates the minimum in the dipolar contribution of the spin-wave energy. Since the energy contributions and the demagnetization tensor change with the thickness of the investigated waveguide, the critical width below which the spins become unpinned is different for different waveguide thicknesses. This is shown in Fig. \ref{fig3}(c), where the inverse effective width $w/w_\mathrm{eff}$ is shown for different waveguide thicknesses. Symbols are the results of micromagnetic simulations, lines are calculated semi-analytically. As can be seen from the figure, the critical width linearly increases with increasing thickness. This is summarized in Fig. \ref{fig3}(d), which shows the critical width (i.e. the maximum width for which $w/w_\mathrm{eff} = 0$) as a function of thickness.

The critical widths for YIG, Permalloy, CoFeB and the Heusler compound Co$_2$Mn$_{0.6}$Fe$_{0.4}$Si with different thicknesses are investigated. Figure \ref{fig3}(e) shows the inverse effective width $w/w_\mathrm{eff}$ as a function of the waveguide width for these materials which can be considered as typical materials used in magnonics. Figure \ref{fig3}(f) shows the critical width ($w_\mathrm{crit}$) as a function of the exchange length $\lambda$ for different thicknesses. A simple empirical linear formula is found by fitting the critical widths for different materials in a wide range of thicknesses to estimate the critical width:
\begin{equation}
    \label{e4}
    w_\mathrm{crit} = 2.2 h + 6.7\lambda
\end{equation}
where $h$ is the thickness of the waveguide and $\lambda$ is the exchange length given by $\lambda = \sqrt{2A / (\mu_0M_\mathrm{s}^2)}$  with the exchange constant $A$, the vacuum permeability $\mu_0$, and the saturation magnetization $M_\mathrm{s}$.

\subsection{Profiles of higher-order width modes}
In \cite{Wan19prl}, only the profile of the fundamental mode ($n = 0$) has been discussed, therefore the mode profiles of higher width modes are shown in Fig. \ref{fig4} for the widths of the waveguides of $5\,\mu$m corresponding to the practically fully pinned case (Fig. \ref{fig4}(a)), $50\,$nm representing fully unpinned case  (Fig. \ref{fig4}(c)), and $1\,\mu$m which can be considered as an intermediate case (Fig. \ref{fig4}(b)). For a $5\,\mu$m wide waveguide all higher width modes are clearly partially pinned due to the large width and an insufficient contribution of the exchange energy. In contrast to this the higher modes of a $1\,\mu$m wide waveguide are clearly unpinned for modes $n>2$. Since the fundamental mode is already unpinned for a $50\,$nm wide waveguide also all higher width modes are fully unpinned.

\begin{figure*}[tbh!]
    \vskip1mm
    \includegraphics[width=0.8\textwidth]{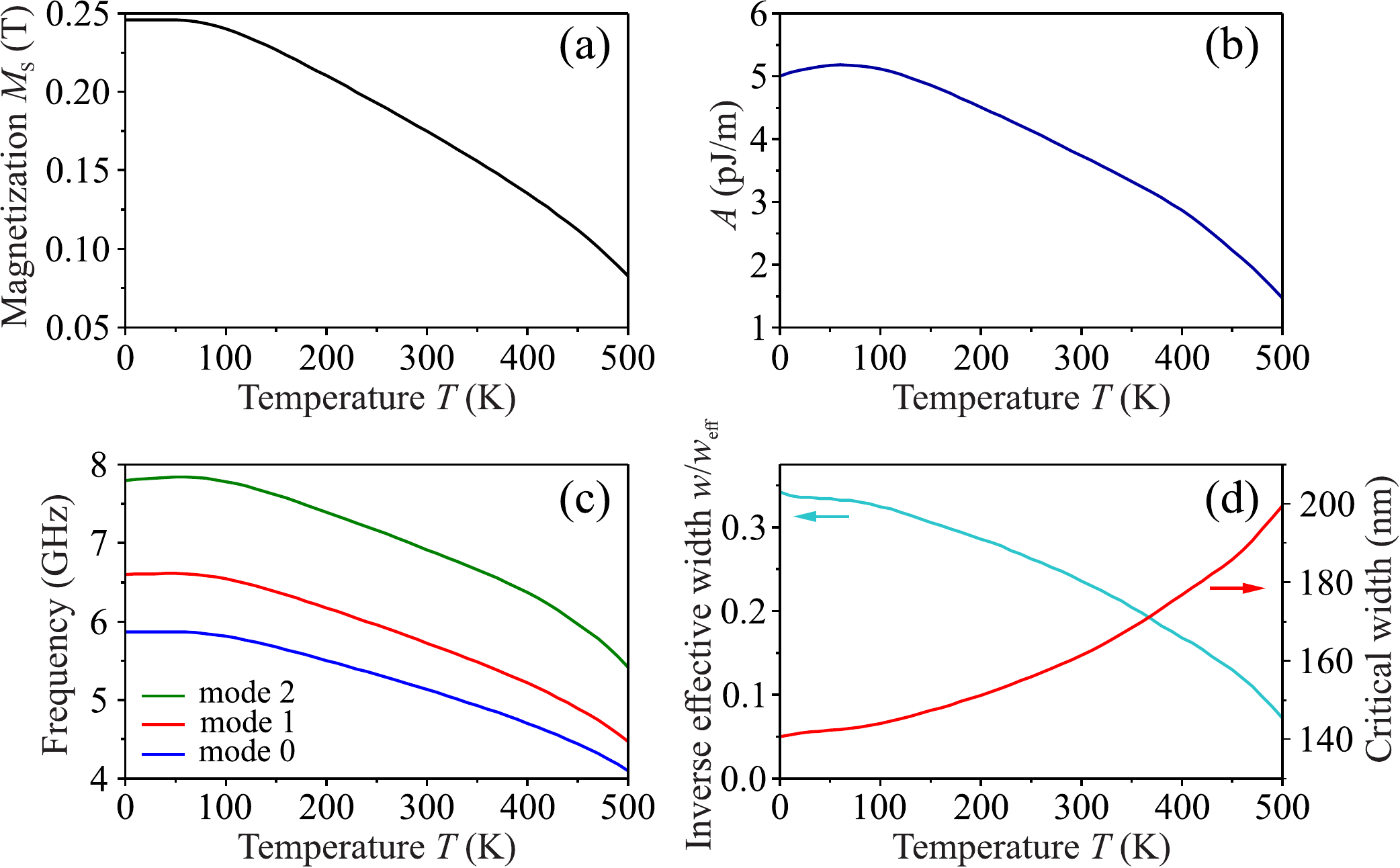}
    \vskip-3mm\caption{Temperature dependence of the saturation magnetization (a) and exchange constant (b) of YIG. (c) The temperature dependencies of the frequencies of the first three modes for a YIG waveguide with $h = 20\,$nm, $w = 200\,$nm and $B_0 = 108.9\,$mT. (d) The temperature dependence of the inverse effective width (left axis) and critical width of the exchange unpinning (right axis).}
    \label{fig5}
\end{figure*}

\subsection{Temperature dependence of spin pinning and frequencies of the spin-wave modes}

In the following, the quasi-analytic theory is used to study the influence of the temperature on the discussed phenomena. There are two main parameters that introduce the temperature dependence of the spin-wave dispersion, the pinning condition and the pinning parameter: the saturation magnetization $M_\mathrm{s}$ and the exchange constant $A$. Furthermore, the temperature dependence of the surface anisotropy constant $K_\textrm{s}$ at the lateral edges of the waveguide, can lead to an additional temperature dependence of the pinning parameter $d$ (see Eq. (\ref{s4})). However, typically this dependency is rather weak and is therefore neglected in the following. The calculated saturation magnetization $M_\mathrm{s}$ for YIG is shown in Fig. \ref{fig5}(a) as a function of the temperature and was obtained using the theoretical model developed in \cite{Han74jap}. The experimentally measured temperature dependence of the exchange constant $A$ taken from \cite{Lec61jap} is shown in Fig. \ref{fig5}(b).

Figure \ref{fig5}(c) shows the resulting temperature dependency of the frequencies of the first three modes for a YIG waveguide of thickness $h = 20\,$nm, width $w = 200\,$nm, and for an external magnetic field $B_0 = 108.9\,$mT applied along the stripe. One can clearly see that the frequencies of all modes decrease with the increase in temperature due to the decrease in the saturation magnetization. The critical width, at which the unpinning takes place, depends on both the saturation magnetization and the exchange constant as it can be seen e.g. from the empirical Eq. (\ref{e4}). The interplay between the both dependencies results in the increase of the critical width with the increase in temperature from the value of around $140\,$nm for zero temperature up to around $200\,$nm for $500\,$K - see Fig. \ref{fig5}(d). At the same time, the spin pinning, which is shown in the same figure in terms of inverse effective width of the waveguide $w/w_\mathrm{eff}$, decreases with the increase in temperature. This happens due to the dominant contribution of the temperature dependence of the saturation magnetization which, consequently, defines the strength of the dipolar pinning phenomenon. To conclude, if one conducts low temperature experiments which rely or require a fully unpinned state of the system a careful design of the structure dimensions is necessary.

\subsection{Spin-wave dispersion in nano-strucutres and the dependence of spin pinning of spin-wave wavenumber}

\begin{figure}[tbh!]
    \vskip1mm
    \includegraphics[width=0.8\columnwidth]{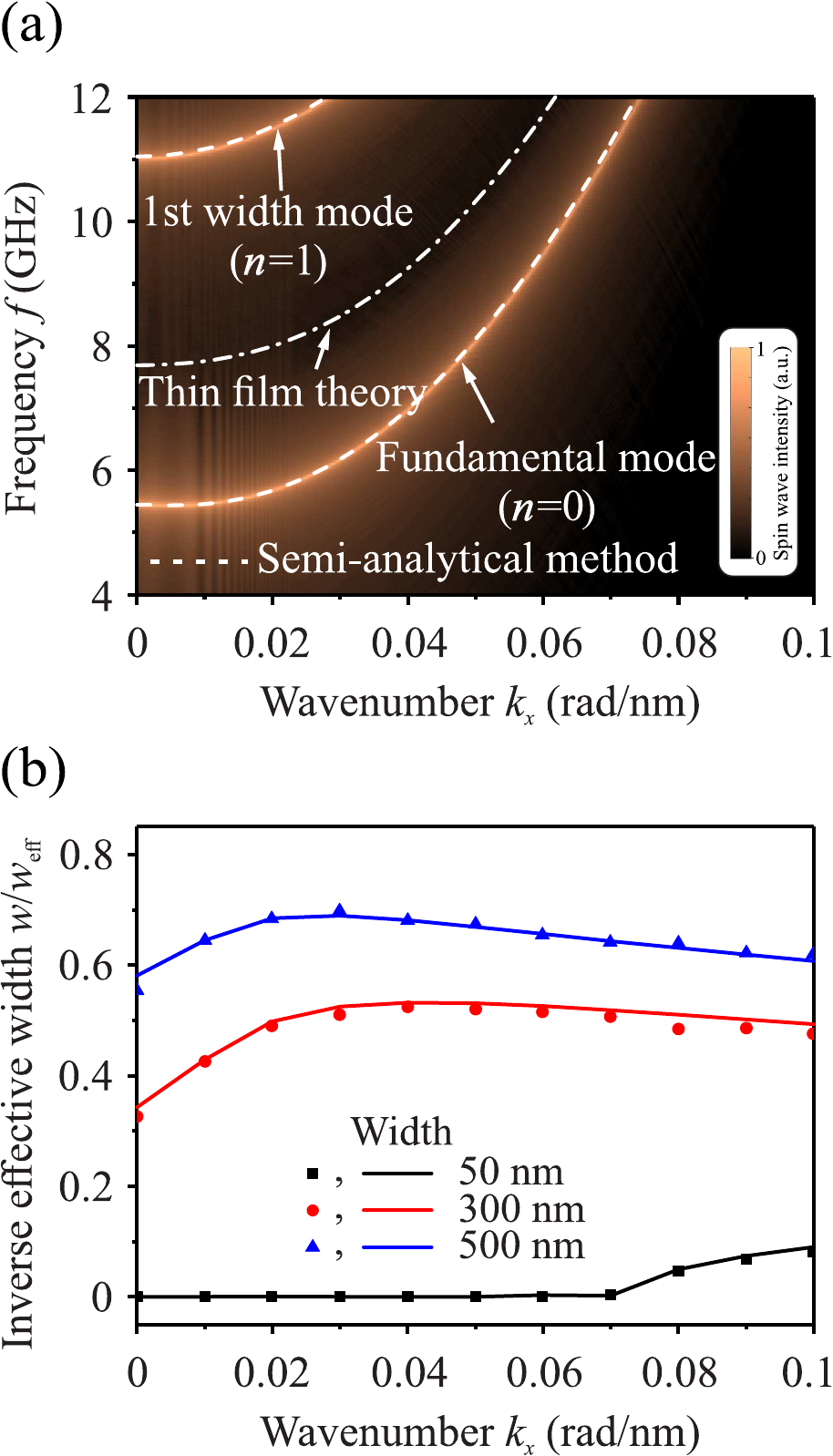}
    \vskip-3mm\caption{(a) Spin-wave dispersion relation of the first two width modes from micromagnetic simulations (color-code) and theory (dashed lines). (b) Inverse effective width $w/w_\mathrm{eff}$ as a function of the spin-wave wavenumber $k_x$ for different thicknesses and waveguide widths, respectively. \cite{Wan19prl} }
    \label{fig6}
\end{figure}

Up to now, the discussion was limited to the special case of $k_x = 0$. In the following, the influence of finite wave vector will be addressed. The spin-wave dispersion relation of the fundamental ($n = 0$) mode obtained from micromagnetic simulations (color-code) together with the semi-analytical solution (white dashed line) are shown in Fig. \ref{fig6}(a) for the YIG waveguide of $w = 50$\,nm width. The figure also shows the low-wavenumber part of the dispersion of the first width mode ($n = 1$), which is pushed up significantly in frequency due to its large exchange contribution. Both modes are described accurately by the quasi-analytical theory. As it is described above, the spins are fully unpinned in this particular case. In order to demonstrate the influence of the pinning conditions on the spin-wave dispersion, a hypothetic dispersion relation for the case of partial pinning is shown in the figure with a dash-dotted white line (the case of $w/w_\mathrm{eff} = 0.63$ is considered that would result from the usage of the thin strip approximation \cite{Gus02prb}). One can clearly see that the spin-wave frequencies in this case are considerably higher. Figure \ref{fig6}(b) shows the inverse effective width $w/w_\mathrm{eff}$ as a function of the wavenumber $k_x$ for three exemplary waveguide widths of $w = 50$\,nm, $300$\,nm and $500$\,nm. As it can be seen, the effective width and, consequently, the ratio $w/w_\mathrm{eff}$ shows only a weak nonmonotonic dependence on the spin-wave wavenumber in the propagation direction. This dependence is a result of an increase of the inhomogeneity of the dipolar fields near the edges for larger $k_x$, which increases pinning \cite{Gus11mmm}, and of the simultaneous decrease of the overall strength of dynamic dipolar fields for shorter spin waves. Please note that the mode profiles are not only important for the spin-wave dispersion. The unpinned mode profiles also greatly improve the coupling efficiency between two adjacent waveguides \cite{Wan18sca,Sad15apl,Sad18prl,Sad17prb}.

\subsection{Spin-wave lifetime in magnetic nanostructures}

\begin{figure}[tbh!]
    \vskip1mm
    \includegraphics[width=0.9\columnwidth]{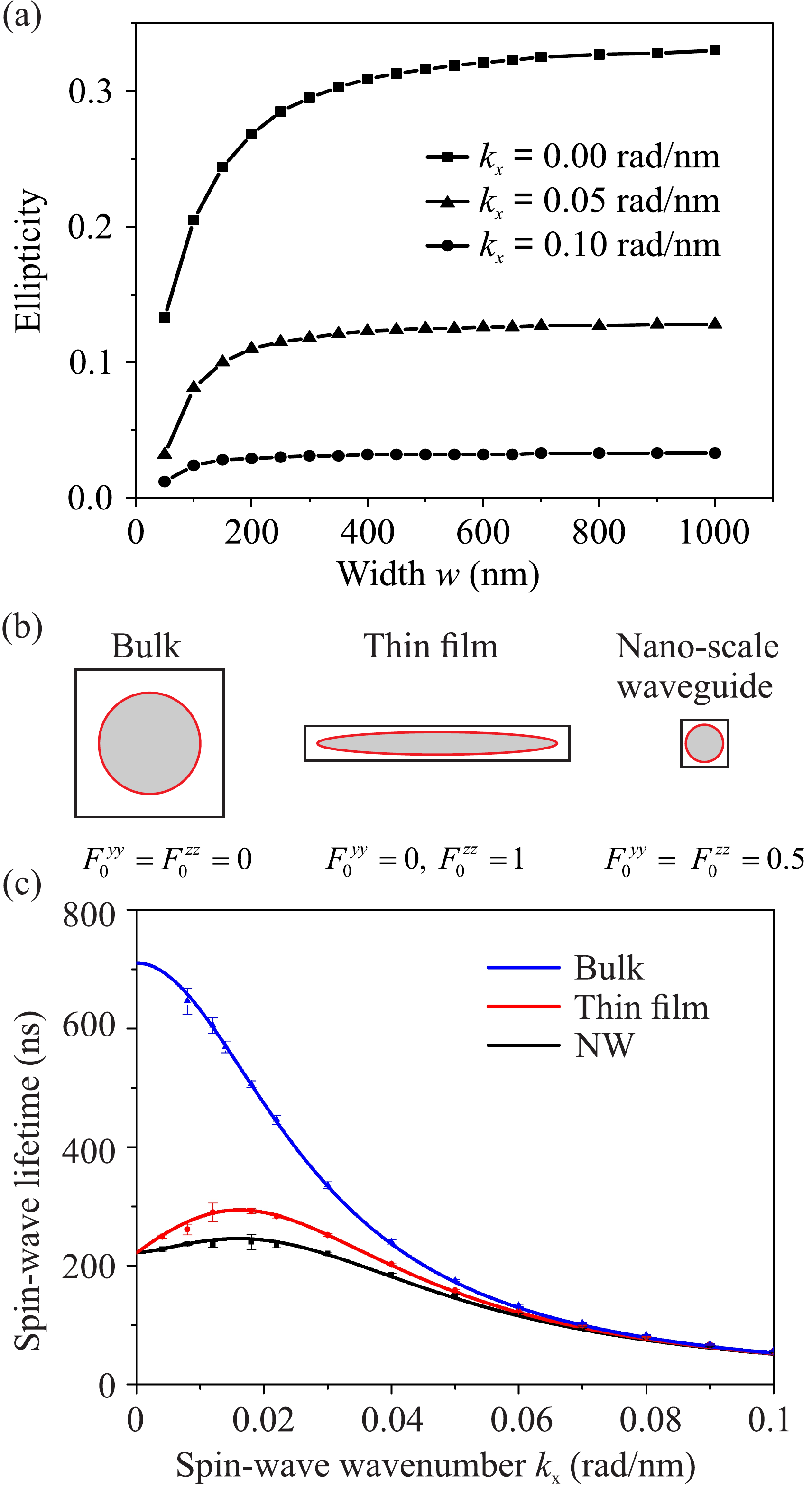}
    \vskip-3mm\caption{(a) Ellipticity as a function of the waveguide width for different spin-wave wavenumbers $k_x$ for a thickness of $h = 39\,$nm and an external magnetic field of $B_0 = 108.9\,$mT (b) The spin precession trajectories (red lines) and the components of the demagnetization tensor $F_0^{yy}$ and $F_0^{zz}$ for different sample geometries. (c) The spin-wave lifetime as a function of the spin-wave wavenumber. The lines and dots are obtained from Eq. (\ref{s10}) and micromagnetic simulation, respectively.}
    \label{fig7}
\end{figure}

The spin-wave lifetime depends on the ellipticity of the magnetization precession, and, thus, on the spin pinning conditions.  The top panel of Fig. \ref{fig2}(b) shows an additional feature of the narrow waveguide: as the aspect ratio of the waveguides approaches unity, the ellipticity of the precession, a well-known feature of micron-sized waveguides which still resemble a thin film \cite{Bra17phr,Gur96boo}, vanishes and the precession becomes nearly circular. In addition, in nanoscale waveguides, the ellipticity is constant across the width, while it can be different at the waveguide center and near its edges for a $1\,\mu$m wide waveguide. In general, the definition of the ellipticity $\epsilon$ of the precession is given by the ratio of the precession components as follows:
\begin{equation}
\label{ss11}
\epsilon = 1-\frac{m_\mathrm{min}}{m_\mathrm{max}},
\end{equation}
where $m_\mathrm{min}$ and $m_\mathrm{max}$ denote the respective amplitudes of the smaller and larger component of the precession. Calculating the average relation between the magnetization components $m_y$ and $m_z$ it follows
\begin{equation}
\label{ss10}
\biggl\vert\frac{m_y}{m_z}\biggl\vert = \sqrt{\left(\frac{(\omega_\mathrm{H} + \omega_\mathrm{M}(\lambda^2K^2 + F^{zz}_{k_x})}{(\omega_\mathrm{H} + \omega_\mathrm{M}(\lambda^2K^2 + F^{yy}_{k_x})})\right)}\:,
\end{equation}
from which the ellipticity can be calculated for any width in dependency of the spin-wave wavenumber $k_x$ as it is shown in Fig. \ref{fig7}(a).

The relaxation lifetime $\tau$ of the uniform precession mode in an infinite medium (without inhomogeneous linewidth $\Delta B_0$) is simply defined as $\tau = 1/(\alpha\omega)$, where $\omega$ is the angular frequency of the spin wave and $\alpha$ is the damping. However, the dynamic demagnetizing field has to be taken into account in finite spin-wave waveguide. The lifetime can be found by the phenomenological model \cite{Sta86jap,Sta09boo,Ver18prb}
\begin{equation}
\label{s9}
    \tau = \left(\alpha\omega\frac{\partial \omega}{\partial \omega_\mathrm{H}}\right)^{-1}.
\end{equation}

The dispersion relation has been shown in the manuscript (Eq. (\ref{e3})). The demagnetization tensors are independent of $\omega_\mathrm{H}$. Differentiating Eq. (\ref{e3}) yields the lifetime as
\begin{equation}
\label{s10}
    \tau = \left(\frac{1}{2}\alpha(2\omega_\mathrm{H} + 2\omega_\mathrm{M}\lambda^2K^2 + \omega_\mathrm{M} (F^{zz}_{k_x} + F^{yy}_{k_x}))\right)^{-1}.
\end{equation}
This formula clearly shows that the lifetime of the uniform precession ($k_x=0$) depends only on the sum of the dynamic $yy$ and $zz$ components of demagnetization tensors.

Figure \ref{fig7}(b) shows the cross-section, spin precession trajectory (red line) and the dynamic components of the demagnetization tensors of different sample geometries. The spin precession trajectory changes from elliptic for the thin film ($a_\mathrm{r}\ll1$) to circular for the nanoscopic waveguide ($a_\mathrm{r}=1$). The spin precession trajectory in the bulk material is also circular (in the geometry when spin waves propagate parallel to the static magnetic field, the same geometry as studied for nanoscale waveguides).
The dependence of the lifetime on the wavenumber is shown in Fig. \ref{fig7}(c) for YIG with a damping constant   $\alpha = 2\times 10^{-4}$. The inhomogeneous linewidth is not taken into account. The lifetime of the uniform precession ($k_x=0$) for the bulk material is much large than that in the thin film and nanoscopic waveguide, another consequence of the absence of dynamic demagnetization in the bulk ($F^{zz}_{0} = F^{yy}_0$). Moreover, the lifetimes of the uniform precession ($k_x=0$) for a thin film (red line) and for a nanoscopic waveguide (black line) have the same value, because the lifetime depends only on the sum of the two components, which is the same for both cases.

Moreover, the $yy$ and $zz$ components of the demagnetization tensor decrease with an increase of the spin-wave wavenumber (instead, the $xx$ component, which does not affect the spin wave dynamic in our geometry, increases). The lifetime is inversely proportional to the square of the wavenumber and the sum of the dynamic demagnetization components. In the exchange region, the lifetime is, thus, dominated by the wavenumber. Therefore, the lifetimes for short-wave spin-waves are nearly the same for the three different geometries.

\begin{figure}[tbh!]
    \vskip1mm
    \includegraphics[width=0.9\columnwidth]{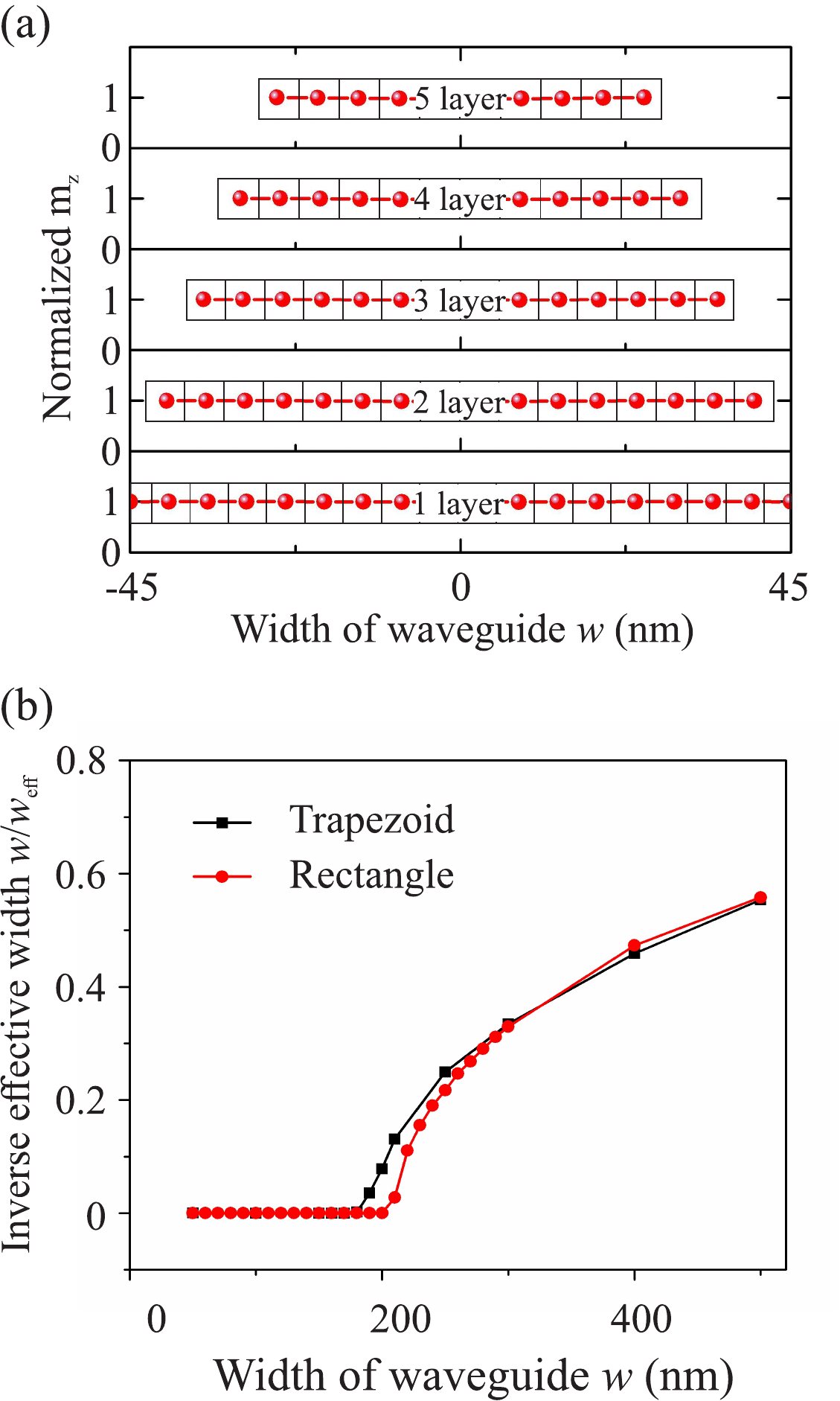}
    \vskip-3mm\caption{(a) Trapezoidal cross section of the simulated waveguide with the normalized spin-wave profile for the different layers. (b) The inverse effective width $w/w_\mathrm{eff}$ as a function of the width of the waveguide for trapezoidal and rectangular form.}
    \label{fig8}
\end{figure}

\subsection{Dependence of the spin pinning on a trapezoidal form of the waveguides}

A perfect rectangular form is not achievable in the experiment due to the involved patterning technique. As a result of the etching, the cross-section of the waveguides is always slightly trapezoidal. In this section, the influence of such a trapezoidal form on the spin pinning conditions is studied. In our experiment, the trapezoidal edges extent for approximately $20$\,nm on both sides for all the patterned waveguides as it can be seen from Fig. \ref{fig1}(d). We performed additional simulation on waveguides with such trapezoidal edges. The simulated cross-section is shown in the top of Fig. \ref{fig8}. The thickness of the waveguide is divided into 5 layers with different widths ranging from $90$\,nm to $50$\,nm. The steps at the edges are hard to be avoided due to the finite difference method used in MuMax3. The spin-wave profiles in the different $z$-layers are shown at the bottom of Fig. \ref{fig8}(a).

The results clearly show that the spin-wave profiles are fully unpinned along the entire thickness. This is due to the fact that the largest width ($90$\,nm) is still far below the critical width. Hence, the influence of the trapezoidal form of the waveguide on the spin pinning condition is negligible for very narrow waveguides. For large waveguides, it also does not have a large impact as the ratio of the edge to the waveguide area becomes close to zero. Quantitatively, the quasi-ferromagnetic resonance frequency in a $50$\,nm wide waveguide decreases from $5.45$\,GHz for the rectangular shape to $5.38$\,GHz for the trapezoidal form due to the increase of the averaged width which, in fact, even closer to the experiment results ($5.35$\,GHz).

The inverse effective width $w/w_\mathrm{eff}$ as a function of the width of the waveguides is simulated for a trapezoidal and a rectangular form and the result is shown in Fig. \ref{fig8}(b). Here, the width is defined by the minimal width for the trapezoidal form, i.e., the width of the top layer. In the case of trapezoidal form, the inverse effective width is averaged over all $5$ layers. The critical width slightly decreases from $200$ nm for the rectangular cross-section to $180$ nm for the trapezoidal form due to the increase of the averaged width. The difference between the inverse effective widths decreases with increasing width of the waveguide and vanishes when the width is larger than $300$\,nm.

Furthermore, it should be noted that the results of the multilayer simulations demonstrate that the assumption of a uniform dynamic magnetization distribution across the thickness that is used in our analytical theory and micromagnetic simulations featuring only one cell in the z dimension is valid.

\subsection{Influence of edge roughness on the spin pinning}

\begin{figure}[t!]
    \vskip1mm
    \includegraphics[width=0.9\columnwidth]{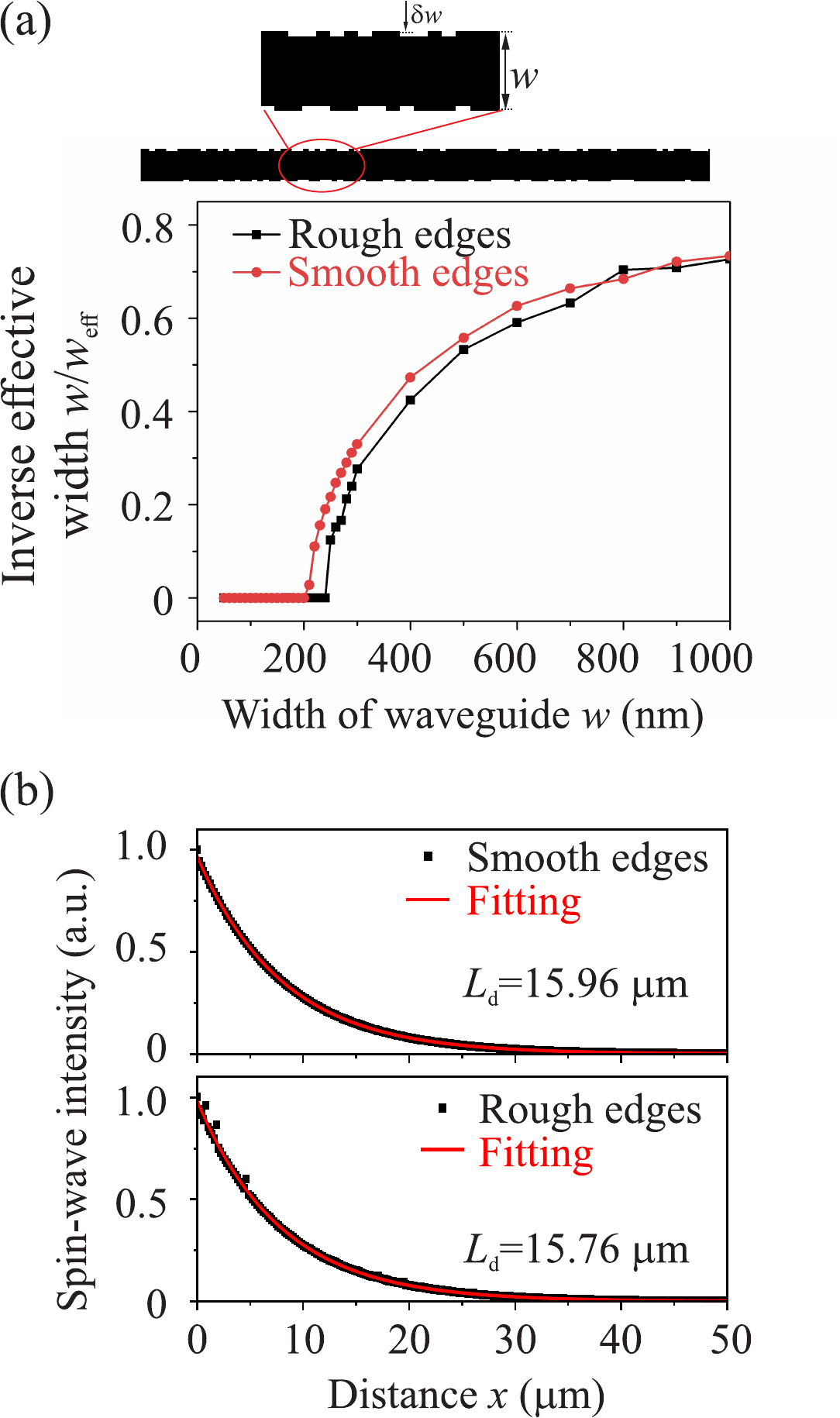}
    \vskip-3mm\caption{(a) Top: Schematic of the rough waveguide and close-up image of the introduced edge roughness. A single randomized defect pattern is generated for each structure width. Bottom: Inverse effective width $w/w_\mathrm{eff}$ as a function of the waveguide width for rough and smooth edges. (b) The normalized spin-wave intensity as a function of the propagation length for $50$\,nm wide waveguides with smooth and rough edges.}
    \label{fig9}
\end{figure}

Perfectly smooth edges are also hard to obtain in the experiment. Therefore, we have considered the influence of edge roughness on the spin pinning. We performed additional simulations on waveguides with rough boundaries for a fixed thickness of $39$\,nm. $5\,$nm (for $50$\,nm to $100$\,nm wide waveguides) or $10$\,nm (for $100$\,nm to $1000$\,nm wide waveguides) wide rectangular nonmagnetic regions with a random length are introduced randomly on both sides of the waveguides to act as defects. The introduction of roughness results in a slight increase of the critical width from $200$\,nm to $240$\,nm, as is shown in Fig. \ref{fig9}(a). These results demonstrate that edge roughness does not have a large influence on spin pinning condition.

Additional simulations are performed to study the influence of a rough edge on the propagation length of spin waves with frequency $6.16$\,GHz ($k_x=0.03$\,rad/nm). Figure \ref{fig9}(b) shows the normalized spin-wave intensity as a function of propagation length for smooth and rough edged waveguide of $450$\,nm width. The decay length slightly decreases from $15.96\,\mu$m for smooth edges to $15.76\,\mu$m for rough edges. Since the spins in nanoscopic waveguides are already unpinned, the effect of such an edge roughness is not too important anymore and the propagation length is essentially unaffected.

\section{Conclusions}
To conclude, an in-detail investigation of the pinning phenomenon based on the theoretical description of \cite{Wan19prl} is presented and the quasi-analytical model is outlined. The dependency of the effective width on the thickness and the material of choice is analysed and a simple empirical formula is found to predict the critical width for a given system. In addition to \cite{Wan19prl}, higher order width modes up to $n=2$ are analysed. An investigation of the effective width for finite wavevectors along the waveguide yields only a weak nonmonotonic dependence. It is shown, that assuming a more realistic trapezoidal cross section of the structures rather than the ideal rectangular shape results in a small decrease of the quasi-FMR frequency and a slight reduction of the critical width. Moreover, the influence of edge roughness is studied which shows a small increase of the critical width compared to the case of smooth edges. Here, also the impact on the decay length of propagating waves is investigated and only a small reduction is found. The temperature dependence of the pinning phenomenon shows that the dependencies of the saturation magnetization and exchange constant of YIG result in the decrease of spin pinning with the increase in temperature and in the increase in the critical width of the exchange unpinning. This assumes that low temperatures are favourable for the dipolar pinning and the sizes of the structures have to be decreased further in order to operate with fully-unpinned uniform spin-wave modes. 

The presented results provide valuable guidelines for applications in nano-magnonics where spin waves propagate in nanoscopic waveguides with aspect ratios close to one and lateral sizes comparable to the sizes of modern CMOS technology.

\vskip3mm \textit{Acknowledgement.}
The authors thank Burkard Hillebrands and Andrei Slavin for valuable discussions. This research has been supported by ERC Starting Grant 678309 MagnonCircuits and by the DFG through the Collaborative Research Center SFB/TRR-173 ``Spin+X'' (Projects B01) and through the Project DU 1427/2-1. B.H. acknowledges support by the Graduate School Material Science in Mainz (MAINZ). R. V. acknowledges support by the National Academy of Sciences of Ukraine Grant No. 23-04/01-2019.

\vskip3mm $^*$These authors have contributed equally to this work.
\vskip3mm $^{**}$\textit{chumak@physik.uni-kl.de}

\bibliographystyle{Style}
\bibliography{./magnonic}

\end{document}